\documentclass[12pt]{iopart}
\usepackage{iopams}  
\usepackage{combelow}
\usepackage{graphicx}
\expandafter\let\csname equation*\endcsname\relax
\expandafter\let\csname endequation*\endcsname\relax
\usepackage{amsmath}
\usepackage{mathtools}
\usepackage{xcolor}
\usepackage{todonotes}
\usepackage{caption}
\usepackage{subcaption}

\definecolor{amaranth}{rgb}{0.9, 0.17, 0.31}

\begin{document}

% \title[\textcolor{amaranth}{A sensitivity-driven sparse grid framework for discharge optimization}]{Turbulence suppression by energetic particles: \textcolor{amaranth}{A sensitivity-driven dimension-adaptive sparse grid framework for discharge optimization}}
\title[A sensitivity-driven sparse grid framework for discharge optimization]{Turbulence suppression by energetic particles: A sensitivity-driven dimension-adaptive sparse grid framework for discharge optimization}

\author{Ionu\cb{t}-Gabriel Farca\cb{s}$^1$, Alessandro Di Siena$^1$, Frank Jenko$^{1, 2, 3}$}
\address{$^1$Oden Institute for Computational Engineering and Sciences, The University of Texas at Austin, 78712 Austin, TX, United States \\
$^2$Max Planck Institute for Plasma Physics, 85748 Garching, Germany \\
$^3$Technical University of Munich, 85748 Garching, Germany}
\ead{ionut.farcas@austin.utexas.edu}

\vspace{10pt}
% \begin{indented}
% \item[]November 2020
% \end{indented}

\vspace{2pc}
\submitto{\NF}
%\maketitle
% \ioptwocol
%

\begin{abstract}

A newly developed sensitivity-driven approach is employed to study the role of energetic particles in suppressing turbulence-inducing micro-instabilities for a set of realistic JET-like cases with NBI deuterium and ICRH $^3$He fast ions. First, the efficiency of the sensitivity-driven approach is showcased for scans in a $21$-dimensional parameter space, for which only $250$ simulations are necessary. The same scan performed with traditional Cartesian grids with only two points in each of the $21$ dimensions would require $2^{21} = 2,097,152$ simulations. Then, a $14$-dimensional parameter subspace is considered, using the sensitivity-driven approach to find an approximation of the parameter-to-growth rate map averaged over nine bi-normal wave-numbers, indicating pathways towards turbulence suppression. The respective turbulent fluxes, obtained via nonlinear simulations for the optimized set of parameters, are reduced by more than two order of magnitude compared to the reference results.

\end{abstract}
\section{Introduction} \label{sec:intro}

\subsection{Turbulence suppression by energetic particles}

One of the most prominent goals of fusion research is the identification of experimental actuators able to suppress cross-field turbulent transport in magnetic confinement devices. Plasma turbulence, which is the main cause for energy confinement degradation, is driven by micro-instabilities due to the unavoidable steep temperature and density gradients.
Among the different paths to suppress turbulent transport which have been explored to date, supra-thermal particles -- generated via auxiliary heating schemes -- have received growing attention recently.  
Observations in both experiments \cite{Bock_NF_2017,Mantica_PRL2009,Mantica_PRL2011,Bonanomi_NF_2018} and numerical simulations \cite{Citrin_PRL,Garcia_NF2015,Doerk_NF_2017,DiSiena_2018,DiSiena_NF_2019,MRomanelli_PPCF2010} have sometimes shown a substantial improvement of plasma confinement in the presence of significant external heating. 
These findings motivate an increasing effort aimed at understanding the physics of the underlying mechanisms responsible for such turbulence stabilization. 
Regarding ion-temperature-gradient (ITG) turbulence, one of the most effective ways of achieving that goal turns out to be the wave-particle resonance interaction discussed in Refs.~\cite{DiSiena_2018,DiSiena_PoP2019,W7X}. 
Here, fast ions are found to interact with an ITG instability when the drift-frequency of the supra-thermal particles gets close to the mode frequency, allowing for an efficient redistribution of free energy. 
The direction and magnitude of the latter might, in principle, depend on a large number of plasma parameters. Therefore, theoretical models are required to provide qualitative and quantitative insights into this process.

A first-principles based model able to capture the essence of this resonance mechanism was derived in Refs.~\cite{DiSiena_2018,DiSiena_PoP2019}, helping to identify the key parameters controlling the underlying physics. 
This model revealed the essentially quasi-linear nature of the basic mechanism and has proven to be remarkably accurate in reproducing and predicting the energetic particle effects on turbulence in both simulations and experiments. 
A good example is provided in Ref.~\cite{F-ATB}, where this reduced model has been applied to design a specific ASDEX Upgrade experiment. 
The observed confinement improvement can be attributed to a new kind of transport barrier, confirming the theoretical predictions. 
Moreover, it has recently been proposed that the wave-particle resonance mechanism should also be able to reduce turbulent transport in optimized stellarators \cite{W7X}.

\subsection{Efficient high-dimensional parameter scans}

Based on these findings, one might want to explore the idea to systematically exploit the potentially strongly stabilizing influence of energetic particles on plasma turbulence. To this aim, it is necessary, however, to study the influence of a large number of parameters which can be used to characterize the properties of the magnetic geometry, the bulk species, and the energetic particles. Addressing the challenge to efficiently carry out scans in high-dimensional parameter spaces will be at the heart of the present paper. In this context, we will also provide an accurate and computationally cheap surrogate model which captures the fast-ion parameter dependencies and which can readily be used for discharge optimization.

Traditionally, parameter scans are performed using full tensor grids, that is, Cartesian products of (equidistant) points.
For example, in Refs.~\cite{Goe14, To13}, a simple uniform scan was carried out, using both linear and nonlinear gyrokinetic simulations. 
The focus in both of these studies was on assessing the sensitivity with respect to changes in (only) one of the key parameters, namely the ion temperature gradient. 
Since the number of points in a full tensor grid increases exponentially with the number of parameters, the practicality of standard parameter scans is limited to settings involving only a small to moderate number of parameters.
The exponential growth of full tensor grids is known as the 'curse of dimensionality' in the computational science community.

In recent years, we have witnessed the emergence of several promising alternatives to full tensor grids.
For example, Ref.~\cite{VH18} used nonintrusive stochastic collocation methods based on non-uniform interpolation sequences, prominent in the uncertainty quantification community, in a drift-wave turbulence study from the CSDX linear plasma experiment. Moreover, sparse grid techniques~\cite{BG04} - an established way of overcoming certain limitations of full tensor grid-based strategies - 
have been considered in the fusion community as well, but mainly restricted to test bed scenarios.
In Ref.~\cite{Ya20}, sparse grids were employed to compute the runaway probability of electrons, and in Ref.~\cite{Ko16}, the sparse grid combination technique was employed in linear flux-tube gyrokinetic simulations based on the {\sc Gene} code \cite{Je00}.

\subsection{Sensitivity-driven dimension-adaptive sparse grid interpolation}

To address the practical limitations of full tensor grids, in Refs.~\cite{Fa20, Fa20b}, we proposed an efficient alternative approach called sensitivity-driven dimension-adaptive sparse grid interpolation.
The new approach is based on adaptive sparse grids and Lagrange interpolation.
Using sparse grids instead of full grids for parameter scans leads to tremendous reductions of the number of required simulations.
For example, in a 5D parameter scan, using eight points in each direction leads to a full grid consisting of $8^5 = 32,768$ points in total.
In contrast, a static sparse grid constructed as in Refs.~\cite{Fa20, Fa20b} uses only $792$ points.

The cardinality of a static sparse grid can be further decreased by using adaptivity based on suitably chosen a posteriori refinement indicators.
The strategy of Refs.~\cite{Fa20, Fa20b} is to employ a greedy adaptive algorithm which exploits the fact that in most real-world problems (i) the underlying parameters are anisotropically coupled and (ii) only a subset of these parameters are important for the scan.
To ascertain the importance and isotropy of the parameters in the scan, sensitivity information about the parameters is employed to ensure that the directions corresponding to parameters with the large sensitivity are preferentially refined.
Thus, instead of using an isotropic, a priori chosen full grid as in standard parameter scans, our algorithm is based on an a posteriori constructed sparse grid which exploits the anisotropic sensitivities of the underlying parameters.

\subsection{Discharge optimization via high-dimensional parameter scans}

Innovative approaches to carrying out high-dimensional parameter scans take center stage when it comes to theory-driven optimization strategies for plasma discharges in fusion devices. A classic example is the goal to control turbulent transport by energetic particles, as discussed above. Our sensitivity-driven approach addresses this problem by creating an accurate interpolation approximation of the parameter-to-output map which can be used in lieu of the high-fidelity model in optimization procedures.

In this context, our main contribution in the present paper is to apply the newly developed sensitivity-driven dimension-adaptive sparse grid interpolation approach \cite{Fa20, Fa20b} to two important tasks.
First, we use it to perform efficient parameter scans.
For example, in Section \ref{sec:lin_res}, we employ the sensitivity-driven approach in a $21$ parameter scan at a cost of only $250$ numerical simulations.
At the end of the scanning procedure, the sensitivity-driven approach allows a detailed description of the sensitivity of all parameters in the scan as well as of their interactions.
Second, we employ the sensitivity-driven algorithm to find accurate approximations of the input-to-growth rate map which is further used to find the parameter set that minimizes the growth rate at several perpendicular wave-numbers. 
In Section \ref{subsec:nonlin_14D}, we find such surrogates at nine perpendicular wave-numbers for at most $230$ simulations per wave-number in a setting with $14$ parameters. 

\subsection{Outline of this paper}

This paper is organized as follows. 
In Section \ref{sec:sens_driven}, we summarize the sensitivity-driven dimension-adaptive sparse grid interpolation approach of Refs.~\cite{Fa20, Fa20b}.
In Section \ref{sec:lin_res}, the sensitivity-driven approach is employed to perform a scan in a $21$ parameter setting, at one perpendicular wave-number.
Here, we demonstrate the detailed description of the sensitivity of all parameters and of their interaction provided by our approach.
In Section \ref{sec:nonlin_res}, we employ the novel method to find surrogates of the parameter-to-growth rate map.
We employ these surrogates to minimize the average growth rate at the most unstable wave-numbers and demonstrate the stabilization provided by the minimizer in linear and nonlinear simulations.
Conclusions are drawn in Section \ref{sec:conclusions}.
\section{An efficient approach for parameter scanning: Sensitivity-driven dimension-adaptive sparse grid interpolation} \label{sec:sens_driven}

In this section, we present the sensitivity-driven dimension-adaptive sparse grid algorithm originally proposed in Refs.~\cite{Fa20,Fa20b} as a basis for our discharge optimization approach.
Our motivation for employing this algorithm is two-fold: (i) we want to efficiently perform parameter scans for a large number of parameters (we will consider scans with $14$ and $21$ parameters) and (ii) we aim to efficiently perform optimization to find the parameter set that minimizes the average growth rate at several perpendicular wave-numbers.
In the following, our notation is similar to that in Refs.~\cite{Fa20,Fa20b}.

Let $\gamma : \mathcal{X} \rightarrow \mathcal{Y}$ denote the growth rate of the dominant eigenmode, where $\mathcal{X} \subset \mathbb{R}^d$ is the set of $d$ input parameters and $\mathcal{Y} \subset \mathbb{R}$.
The growth rate, $\gamma$, depends on $d$ parameters $\boldsymbol{\theta} := (\theta_1, \theta_2, \ldots, \theta_d) \in \mathcal{X}$.
We note that the presented algorithm is applicable to any problem in which the underlying function depending on $\boldsymbol{\theta}$ is at least continuous.

The sparse grid interpolation of $\gamma$ reads
\begin{equation} \label{eq:tensor_delta_finite}
\mathcal{U}^{d}_{\mathcal{L}}[\gamma] = \sum_{\boldsymbol{\ell} \in \mathcal{L}} \Delta^{d}_{\boldsymbol{\ell}}[\gamma],
\end{equation}
where $\boldsymbol{\ell} = (\ell_1, \ell_2, \ldots \ell_{d}) \in \mathbb{N}^{d}$ denotes a \emph{multiindex}, $\mathcal{L} \subset \mathbb{N}^d$ is a \emph{multiindex set}, and
\begin{equation} \label{eq:hierarch_surplus}
\Delta^{d}_{\boldsymbol{\ell}}[\gamma] = \sum_{\boldsymbol{z} \in \{0, 1\}^{d}} (-1)^{|\boldsymbol{z}|_1} \mathcal{U}^{d}_{\boldsymbol{\ell}}[\gamma]
\end{equation}
are the so-called \emph{hierarchical surpluses}, where $|\boldsymbol{z}|_1 := \sum_{i=1}^{d} z_i$.
The surpluses are computed from full-grid operators, $\mathcal{U}^{d}_{\boldsymbol{\ell}}$, which are tensorizations of $1$D approximations:
\begin{equation} \label{eq:tensor_1D}
\mathcal{U}^{d}_{\boldsymbol{\ell}}[\gamma] = \left(\bigotimes_{i=1}^{d} \mathcal{U}^{i}_{\ell_i}\right)[\gamma].
\end{equation}
Therefore, the sparse grid interpolation approximation Eq.~\eqref{eq:tensor_delta_finite} is a linear combination of hierarchical surpluses Eq.~\eqref{eq:hierarch_surplus}, which are computed from tensorizations Eq.~\eqref{eq:tensor_1D} of one-dimensional operators, $\mathcal{U}^{i}_{\ell_i}$.

To define the sparse grid approximation Eq.~\eqref{eq:tensor_delta_finite}, we need to specify (i) the one-dimensional operators, $\mathcal{U}^{i}_{\ell_i}$, and (ii) the multiindex set, $\mathcal{L}$.
In this work, $\mathcal{U}^{i}_{\ell_i}$ are interpolation operators based on Lagrange polynomials constructed using (L)-Leja interpolation knots.
For improved numerical stability, we implement Lagrange interpolation in terms of the so-called first form of the barycentric formula \cite{BT04}.
The (L)-Leja points are determined greedily as:
\begin{equation} \label{eq:leja}
\begin{split}
& u_1 = 0.5, \\
\ & u_n = \underset{u \in \mathcal{X}_i}{\mathrm{argmax}} \prod_{m=1}^{n-1} \left| (u - u_m) \right|, \quad n = 2, 3, \ldots, \ell_i,
\end{split}
\end{equation}
where $\mathcal{X}_i$ is the domain of parameter $\theta_i$.
We note that by construction, the (L)-Leja sequence Eq.~\eqref{eq:leja} is interpolatory and arbitrarily granular.
In addition, as it was shown in \cite{NJ14}, (L)-Leja points are very accurate for interpolation. 
For more details on (L)-Leja points, see Refs.~\cite{Fa20, Fa20b, NJ14} and the references therein.

To fully define Eq.~\eqref{eq:tensor_delta_finite}, we need to specify the multiindex set, $\mathcal{L}$.
$\mathcal{L}$ is critical for the computational efficiency of constructing the approximation Eq.~\eqref{eq:tensor_delta_finite} %the larger the cardinality of $\mathcal{L}$ is, the larger the cost of finding the approximation. This, in turn, means that the computational cost of performing the parameter scan increases with the cardinality of $\mathcal{L}$.
since the cost of the parameter scan increases with the cardinality of $\mathcal{L}$.

For example, in standard parameter scans, a prescribed number of grid points is employed in each of the $d$ directions.
Let us denote this number by $L_{\mathrm{max}} \in \mathbb{N}$.
Using the notation and concepts introduced in this section, the standard parameter scan corresponds to an approximation Eq.~\eqref{eq:tensor_delta_finite} with a multiindex set
\begin{equation} \label{eq:std_param_scans}
    \mathcal{L} = \{\boldsymbol{\ell} \in \mathbb{N}^d: |\boldsymbol{\ell}|_{\infty} \leq L_{\mathrm{max}}\}.
\end{equation}
In standard parameter scans the multiindex set is thus a priori fixed and has a cardinality in $\mathcal{O}(L_{\mathrm{max}}^d)$ which prohibits this method to setups with a large number of parameters.
In addition, since the same number of points is used in all $d$ directions, standard scans treat the $d$ parameters isotropically.

The sensitivity-driven algorithm, in contrast, exploits the fact that in most problems (i) the underlying parameters are anisotropically coupled and (ii) only a subset of these parameters are important for the scan.
To ascertain the importance and isotropy of the parameters in the scan, we employ sensitivity information in terms of Sobol' indices \cite{So01}.
Scanning parameters within prescribed bounds leads to a statistical description of the output of interest, out of which the typical quantities of interest are its expectation and variance.
The variance can be seen as a measure of uncertainty due to the variation of the input parameters within the prescribed bounds.
A \emph{local Sobol' index} measures the percentage contribution of individual parameters and interactions thereof to the variance.
Therefore, the local Sobol' indices add up to $100 \%$.
In a scan involving $d$ parameters, we have $2^d - 1$ local Sobol' indices in total: first $d$ correspond to the individual parameters, next $d(d-1)/2$ indices are for interactions between pairs of two parameters etc.
Based on the local Sobol' indices, we compute a sensitivity score, $s \in \mathbb{N}$, which drives the adaptive process.
Initially, $s = 0$.
We prescribe $d + 1$ tolerances $\boldsymbol{\tau} := (\tau_1, \tau_2, \ldots, \tau_d, \tau_{d+1})$: one for each individual direction ($d$ in total) and the last one for all interactions.
We compare the $d$ individual local Sobol' indices with the first $d$ tolerances and the summation of all local Sobol' indices corresponding to interactions with the last tolerance, and whenever a tolerance is not exceeded, $s$ is increased by one.
Thus, $s$ takes values between zero and $d + 1$.
A large sensitivity score signals significantly large sensitivities in several directions which therefore necessitate refinement.
If the score is zero, on the other hand, there is no need for more refinement.
This is contrast with standard parameter scans which treat the input parameters isotropically.

At the end of the sensitivity-driven adaptive procedure, we obtain a detailed description about the sensitivities -- both individual and due to interactions -- of the parameters in the scan in terms of local Sobol' indices.
Besides local Sobol' indices, we also compute so-called \emph{total Sobol' indices}.
A total Sobol' index is the summation of all local Sobol' indices corresponding to that input.
Therefore, we have $d$ total Sobol' indices for a scan involving $d$ parameters.
Note that parameter interactions are added to multiple total Sobol' indices.
For example, in a setting with two parameters, both total Sobol' indices comprise the local Sobol' index corresponding to the parameter interaction.
We note that the sensitivity-driven approach yields the expectation and variance of $\gamma$ as well, but our emphasis in this work is only on the sensitivity description in terms of local and total Sobol' indices, which is the most valuable in the present context.
% The summation of the total indices serves as an indicator for how strongly the parameters interact: if the total Sobol' indices sum up to a value close to $1$, then the interactions between the parameters are week.
% In general, in practical applications only few interactions between a small number of parameters are significant.
The detailed information about the underlying parameters given by Sobol' indices is, to the best of our knowledge, a novelty in the context of plasma physics applications.
Finally, since the sensitivity-driven dimension-adaptive approach is based on interpolation, it also provides a surrogate of the input-output map.
This surrogate can be further used in computationally expensive tasks such as optimization which are otherwise prohibitive when employing the high-fidelity simulation code.

Before showing our numerical results, we make the following two important observations.
First, recall that the parameters-to-growth rate mapping needs to be at least continuous for the presented sensitivity-driven approach to be applicable; the smoother this mapping is, the better.
Therefore, our sensitivity-driven approach is applicable even in problems in which the growth rate has multiple local minima or maxima, albeit at a likely higher computational cost.
Put differently, if the underlying problem has structure in terms of lower-intrinsic dimensionality or anisotropic coupling of the parameters in the scan, the sensitivity-driven approach will intrinsically exploit that structure and will likely be computationally cheap.
If, in contrast, the parameters are strongly coupled (which is, however, not typical in high-dimensional real-world problems), the sensitivity-driven approach remains applicable, but at a potentially larger computational cost.
We note, nevertheless, that in the latter situation, most approximation algorithms will fail to outperform full-grid approaches.

Second, the important ingredient in the sensitivity-driven approach are the tolerances $\boldsymbol{\tau}$: if these are too large, our approach might stop prematurely and not explore, for example, multiple local extrema.
In these scenarios, one solution is to decrease the tolerances.
Since the sensitivity-driven approach is hierarchical, decreasing the tolerances allows reusing the results obtained using the larger tolerances.
We note that in our numerical experiments thus far, tolerances between $10^{-6} \cdot \boldsymbol{1}_{d+1}$ and $10^{-8} \cdot \boldsymbol{1}_{d+1}$ were sufficient.

\section{21-dimensional parameter scan with only $250$ {\sc Gene} simulations} \label{sec:lin_res}

In the following, we apply the sensitivity-driven dimension-adaptive sparse grid approach to the study of the role of supra-thermal particles in suppressing ITG instabilities for realistic JET-like cases with both NBI fast deuterium and ICRH $^3$He. 
Carbon impurities are considered as well to mimic JET C-Wall plasma conditions. 
This scenario is inspired by Ref.~\cite{DiSiena_2018}, where the presence of the energetic particle species was shown to strongly reduce the ITG drive via a quasi-linear wave-particle interaction.

\subsection{Simulation setup}

The simulations are performed with the gyrokinetic code {\sc Gene} \cite{Je00} in the flux-tube limit. 
Here, we assume periodic boundary conditions in the radial and bi-normal directions and use a Fourier decomposition of the perturbed quantities in the bi-normal direction \cite{Dannert_2005}.
Given the fundamentally electrostatic nature of the wave-particle resonance mechanism, the electron beta is reduced to $\beta_e = 10^{-5}$.
This assumption ensures that no mode transition can occur between ITG instabilities and electromagnetic modes (such as kinetic ballooning modes or Alfv\'en eigenmodes), which could create a discontinuity in the growth rate function, deleterious to the sensitivity-driven approach, which assumes at least continuity of the target function. 
For simplicity, both thermal and fast ion species have been modeled with an equivalent Maxwellian distribution function. 
Choosing a more realistic background distribution for the supra-thermal particles is not expected to qualitative affect the numerical results, as shown in Ref.~\cite{DiSiena_PoP_2018} for a similar set of plasma parameters. 
Furthermore, the magnetic geometry is described by an analytical Miller equilibrium \cite{Miller_PoP98}.
Finally, the growth rates were computed with three digits of accuracy.

\subsection{Application of the sensitivity-driven sparse grid approach}

We now use the sensitivity-driven approach to compute the local and total Sobol' indices of the growth rate of the dominant eigenmode, using a tolerance $\boldsymbol{\tau} = 10^{-6} \cdot \boldsymbol{1}$.
We will denote by $S_p$ the local Sobol' index corresponding to parameter $p$, by $S_{p_1, p_2, \ldots}$ the local Sobol' index associated to the interaction between parameters $p_1, p_2, \ldots$, and by $S^T_p$ the total Sobol' index corresponding to parameter $p$.

\begin{table}[htbp]
\centering
\begin{tabular}{|c|c|c|c|c|}
\hline
$\boldsymbol{\theta}$ & input parameter & nominal value & left bound & right bound \\ 
\hline
$\theta_1$ & $q$ & $1.7364$ & $1.3023$ & $2.1705$ \\
$\theta_2$ & $\hat{s}$ & $0.5226$ & $0.3920$ & $0.6533$ \\
$\theta_3$ & $\omega_{T_i}$ & $4.5640$ & $3.4230$ & $5.7050$ \\
$\theta_4$ & $\omega_{n_i}$ & $0.0063$ & $0.0047$  & $0.0078$ \\
$\theta_5$ & $T_i/T_e$ & $0.9000$ & $0.6750$ & $1.1250$ \\
$\theta_6$ & $\omega_{T_{fi}}$ & $1.0323$ & $0.5000$ & $3.5000$ \\
$\theta_7$ & $\omega_{n_{fi}}$ & $4.7217$ & $1.0000$ & $9.0000$ \\
$\theta_8$ & $n_{fi}$ & $0.0600$ & $0.0100$ & $0.1000$\\
$\theta_9$ & $T_{fi}$ & $9.8000$ & $1.0000$ & $40.0000$ \\
$\theta_{10}$ & $\omega_{T_{^{3}{He}}}$ & $7.4058$ & $1.0000$ & $15.0000$ \\
$\theta_{11}$ & $\omega_{n_{^{3}{He}}}$ & $0.5027$ & $0.5000$ & $3.5000$ \\
$\theta_{12}$ & $n_{^{3}{He}}$ & $0.0700$ & $0.0100$ & $0.1000$\\
$\theta_{13}$ & $T_{^{3}{He}}$ & $6.9000$ & $1.0000$ & $40.0000$ \\
$\theta_{14}$ & $\omega_{T_{e}}$ & $2.2260$ & $1.6695$ & $2.7824$ \\
$\theta_{15}$ & $drR$ & $-0.1379$ & $-0.1724$ & $-0.1034$ \\
$\theta_{16}$ & $\kappa$ & $1.2646$ & $0.9484$ & $1.5807$ \\
$\theta_{17}$ & $s_{\kappa}$ & $0.0324$ & $0.0227$ & $0.0405$\\
$\theta_{18}$ & $\delta$ & $0.0303$ & $0.0242$ & $0.0379$ \\
$\theta_{19}$ & $s_{\delta}$ & $0.0323$ & $0.0242$ & $0.0404$ \\
$\theta_{20}$ & $\zeta$ & $-0.0003$ & $-0.0004$ & $-0.0002$ \\
$\theta_{21}$ & $s_{\zeta}$ & $-0.0017$ & $-0.0022$ & $-0.0013$\\
\hline
\end{tabular}
% \caption{Summary of the $21$ parameters considered.
% The nominal values are displayed in the third column and they are similar to the parameter set considered in Ref.~\cite{DiSiena_2018,Bravenec_PPCF2016}
% The bounds within which we perform the parameter scan are listed in columns four (left bound) and five (right bound).
% Here, $q$ represents the safety factor, $\hat{s}$ the magnetic shear, $\omega_{n,T}$ the logarithmic density and temperature gradients, $dr R = d R / dr$, $\kappa$ the plasma elongation, $s_\kappa = r d ln(\kappa)/dr$, $\delta$ the triangularity, $s_\delta = r d ln(\delta)/dr$, \textcolor{amaranth}{$\zeta$ the squareness} and $s_\zeta =r d ln(\zeta)/dr$. The Carbon temperature is fixed to the main ion one and its density to the reference value.}
\caption{Summary of the $21$ parameters considered.
The nominal values are displayed in the third column and they are similar to the parameter set considered in Ref.~\cite{DiSiena_2018,Bravenec_PPCF2016}
The bounds within which we perform the parameter scan are listed in columns four (left bound) and five (right bound).
Here, $q$ represents the safety factor, $\hat{s}$ the magnetic shear, $\omega_{n,T}$ the logarithmic density and temperature gradients, $dr R = d R / dr$, $\kappa$ the plasma elongation, $s_\kappa = r d ln(\kappa)/dr$, $\delta$ the triangularity, $s_\delta = r d ln(\delta)/dr$, $\zeta$ the squareness and $s_\zeta =r d ln(\zeta)/dr$. The Carbon temperature is fixed to the main ion one and its density to the reference value.}
\label{tab:all_params}
\end{table}

We start our analyses by scanning over the entire set of $21$ parameters, including variations in the pressure and its gradient for the main ions, electrons, fast deuterium and $^3$He, and magnetic geometry.
These parameters are summarized in Table \ref{tab:all_params}.
The reference values are shown in the third column, while the left and right bounds employed in scanning are displayed in the fourth and last column, respectively.
We consider large bounds for the parameters characterizing the energetic particles.
For the remaining parameters, the bounds are $25 \%$ around the reference value.
For simplicity, we fix the bi-normal mode-number to $k_y \rho_s = 0.5$, which represents one of the most unstable mode numbers in the linear {\sc Gene} simulations. 
Here, $\rho_s = c_s /\Omega_i$ where $c_s = \left(T_e/m_i\right)^{1/2}$ is the sound speed and $\Omega_i$ the ion gyro-frequency.

\begin{figure}[htpb]
\centering
\includegraphics[width=1.0\textwidth]{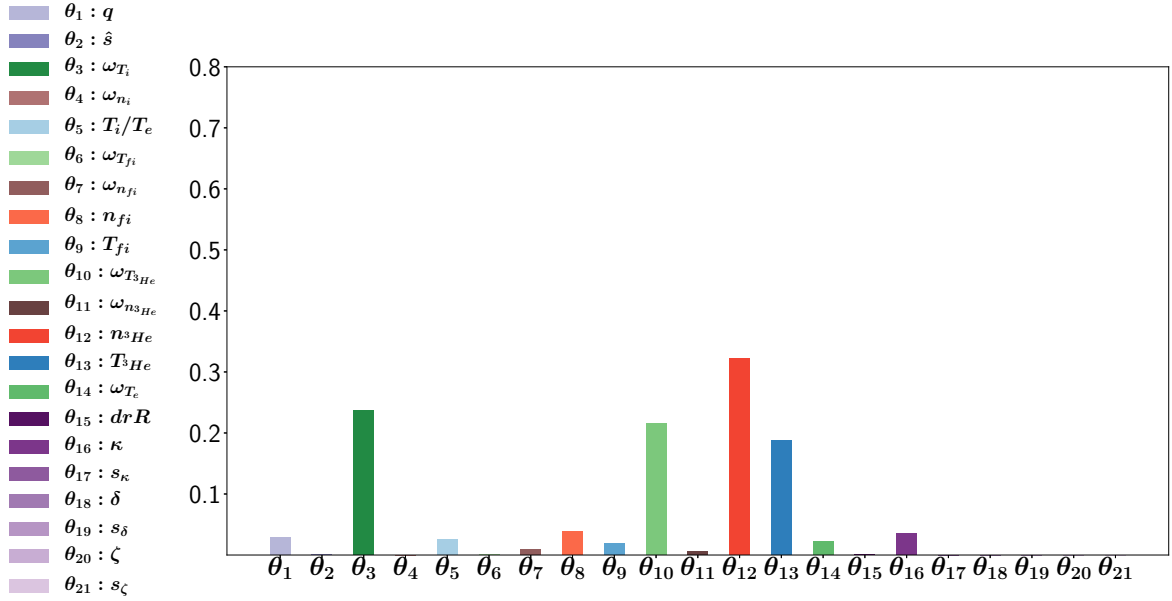}
\caption{Sensitivity information expressed in terms of total Sobol' indices for the scan involving $21$ parameters for $k_y \rho_s = 0.5$.}
\label{fig:sensitivity_test_case_3}
\end{figure}
In Fig.~\ref{fig:sensitivity_test_case_3}, we depict the total Sobol' indices for all $21$ parameters involved in the {\sc Gene} scan. 
The Sobol' indices provide a detailed description of the importance of each of the individual parameters to variations in the linear ITG growth rate, which represents one of the highlights for the sensitivity-driven approach.
By looking at Fig.~\ref{fig:sensitivity_test_case_3}, we observe that the most relevant parameters affecting the ITG growth rates -- corresponding to the larger Sobol' indices -- are: (i) the bulk ion logarithmic temperature gradient, being the main drive of ITG instabilities, (ii) the $^3$He density, (iii) the $^3$He temperature and (iv) the $^3$He logarithmic temperature gradient. %In addition, we note that the geometrical Miller parameters describing the local flux-surface have a negligible impact on the ITG growth rate.

To understand in more detail the role of each of these parameters with respect to the linear ITG instability, we exploit another important feature of the sensitivity-driven sparse grid approach. 
It provides not only a detailed description of the importance of each individual parameters but also of their interactions. 
In other words, we can break down the total Sobol' indices in Fig.~\ref{fig:sensitivity_test_case_3} into contributions of individual inputs and interactions.
In this way, we can identify which parameter interactions are important in the scan, which goes beyond what analyses based on standard parameter scan offer.

\begin{figure}
\begin{subfigure}{0.49\textwidth}
\includegraphics[width=1.0\textwidth]{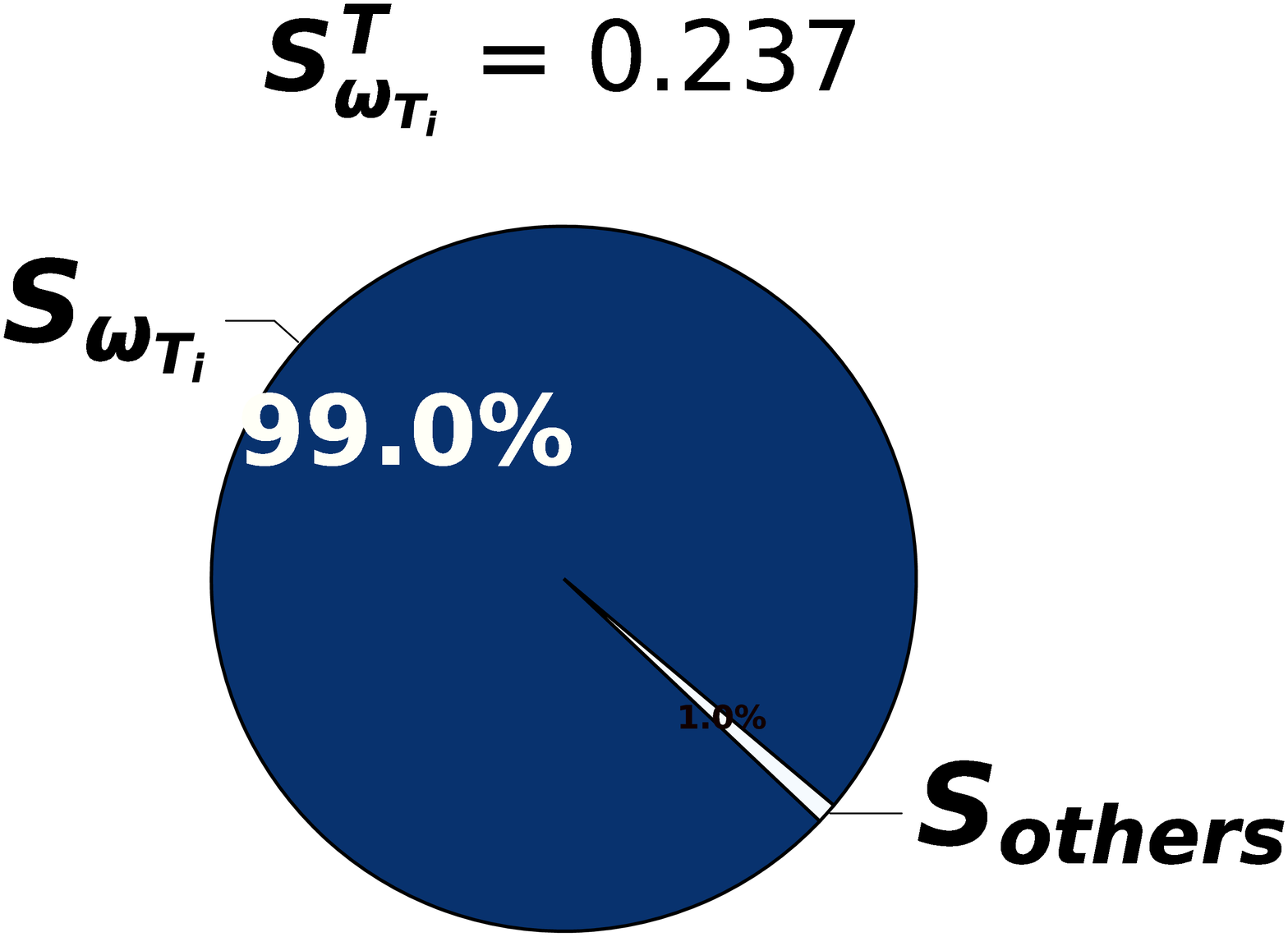}
\caption{Sensitivity analysis for $\omega_{T_i}$}
\end{subfigure}
\begin{subfigure}{0.49\textwidth}
\includegraphics[width=1.0\textwidth]{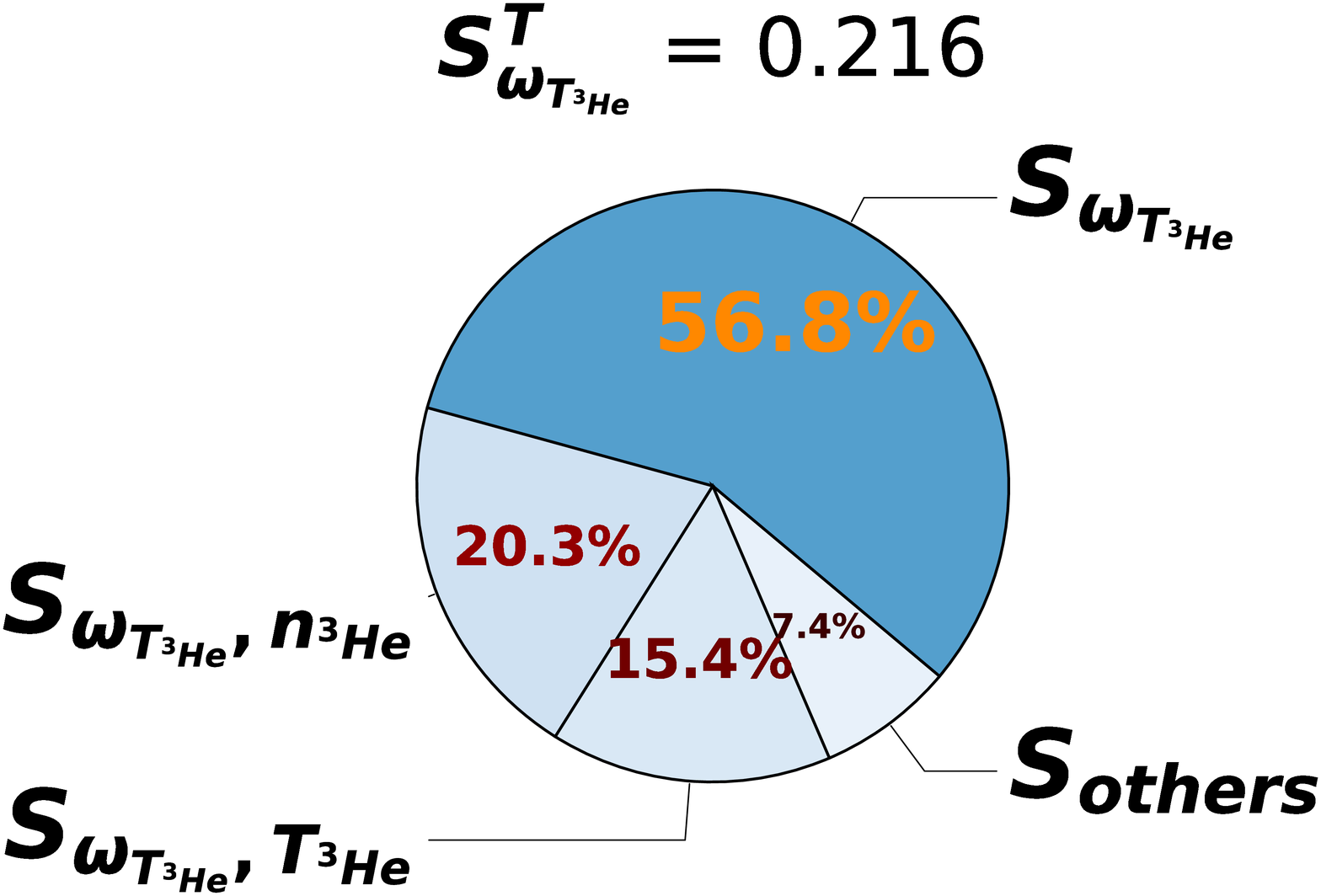}
\caption{Sensitivity analysis for $\omega_{T_{^{3}He}}$}
\end{subfigure}
\begin{subfigure}{0.49\textwidth}
\includegraphics[width=1.0\textwidth]{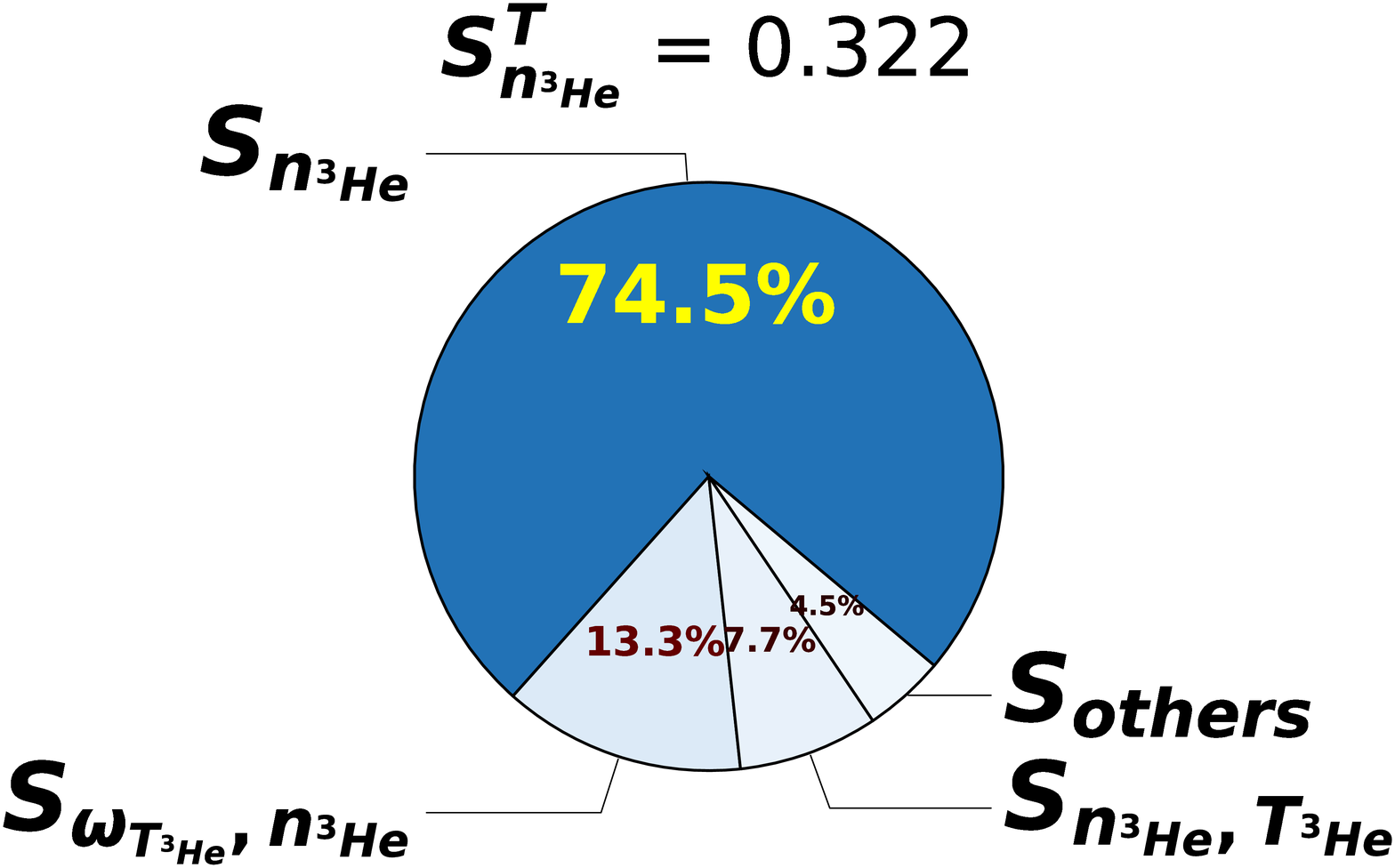}
\caption{Sensitivity analysis for $n_{^{3}He}$}
\end{subfigure}
\begin{subfigure}{0.49\textwidth}
\includegraphics[width=1.0\textwidth]{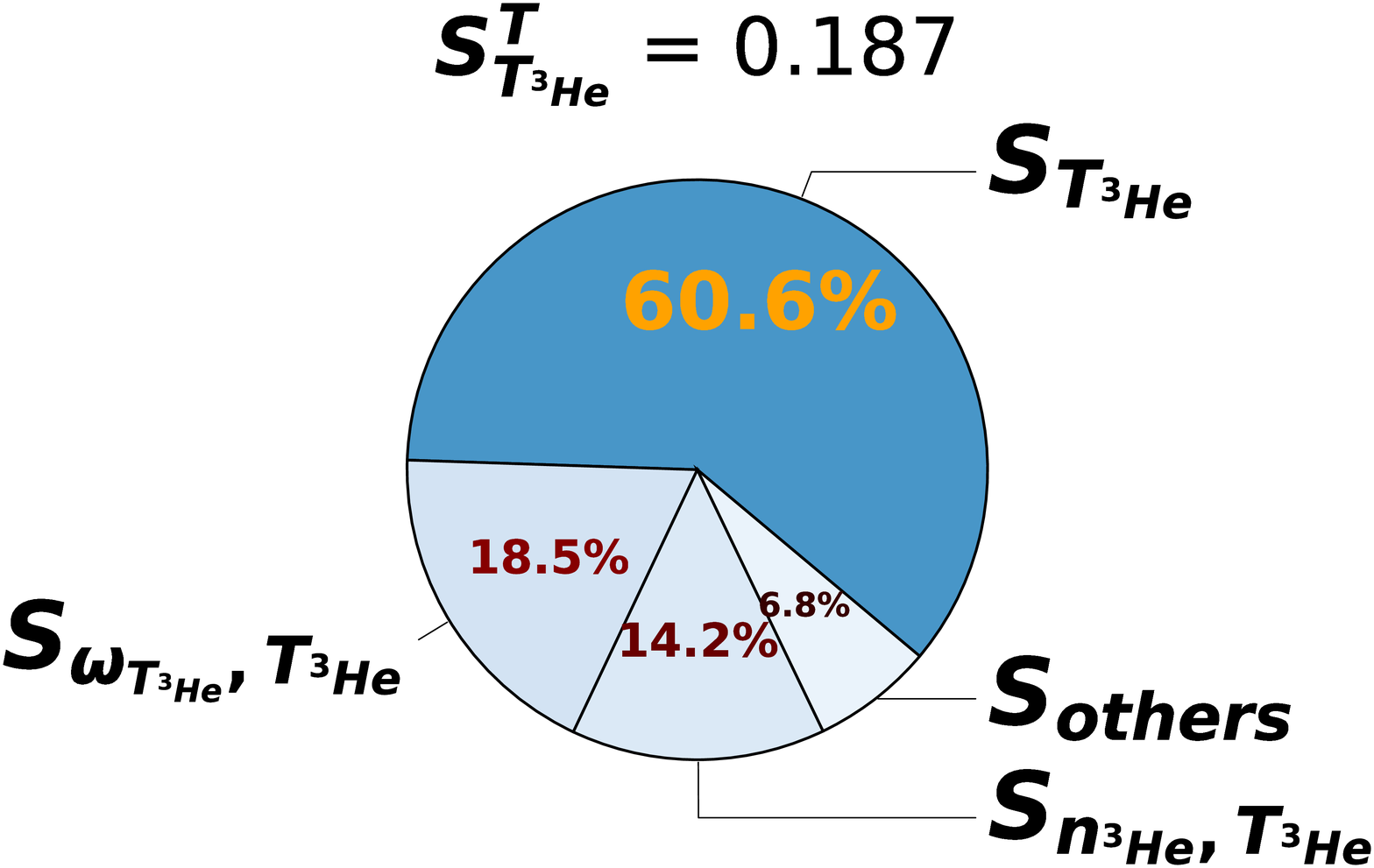}
\caption{Sensitivity analysis for $T_{^{3}He}$}
\end{subfigure}
\caption{Sensitivity due to interactions for (a) $\omega_{T_i}$, (b) $\omega_{T_{^{3}He}}$, (c) $n_{^{3}He}$ and (d) $T_{^{3}He}$ with the other plasma parameters considered.}
\label{fig:pie_charts}
\end{figure}

In Fig.~\ref{fig:pie_charts}, we visualize these contributions for the largest four sensitivity indices (i)-(iv) mentioned above. 
On the top of each figure, we show the value of the corresponding total sensitivity index. 
The pie-charts depict the percentages due to individual parameters and to parameter interactions. 
In addition, the sensitivity indices denoted by $S_{others}$ refer to all remaining local indices that are not visualized in the respective plot.
Fig.~\ref{fig:pie_charts}(a) reveals that the logarithmic main ion temperature gradient does not interact significantly with any of the remaining $20$ parameters. 
Therefore, its large total sensitivity index reflects the large effect that variations of this parameters have on the linear drive of ITG modes. 
In contrast, we observe that the two-parameter interactions involving the $^3$He temperature, its logarithmic gradient and the $^3$He density are significant, accounting for more than $20\%$ of the corresponding total Sobol' index. 
This is shown in Fig.~\ref{fig:pie_charts}(b)--(d).

\begin{figure}
\begin{subfigure}{0.32\textwidth}
\includegraphics[width=1.0\textwidth]{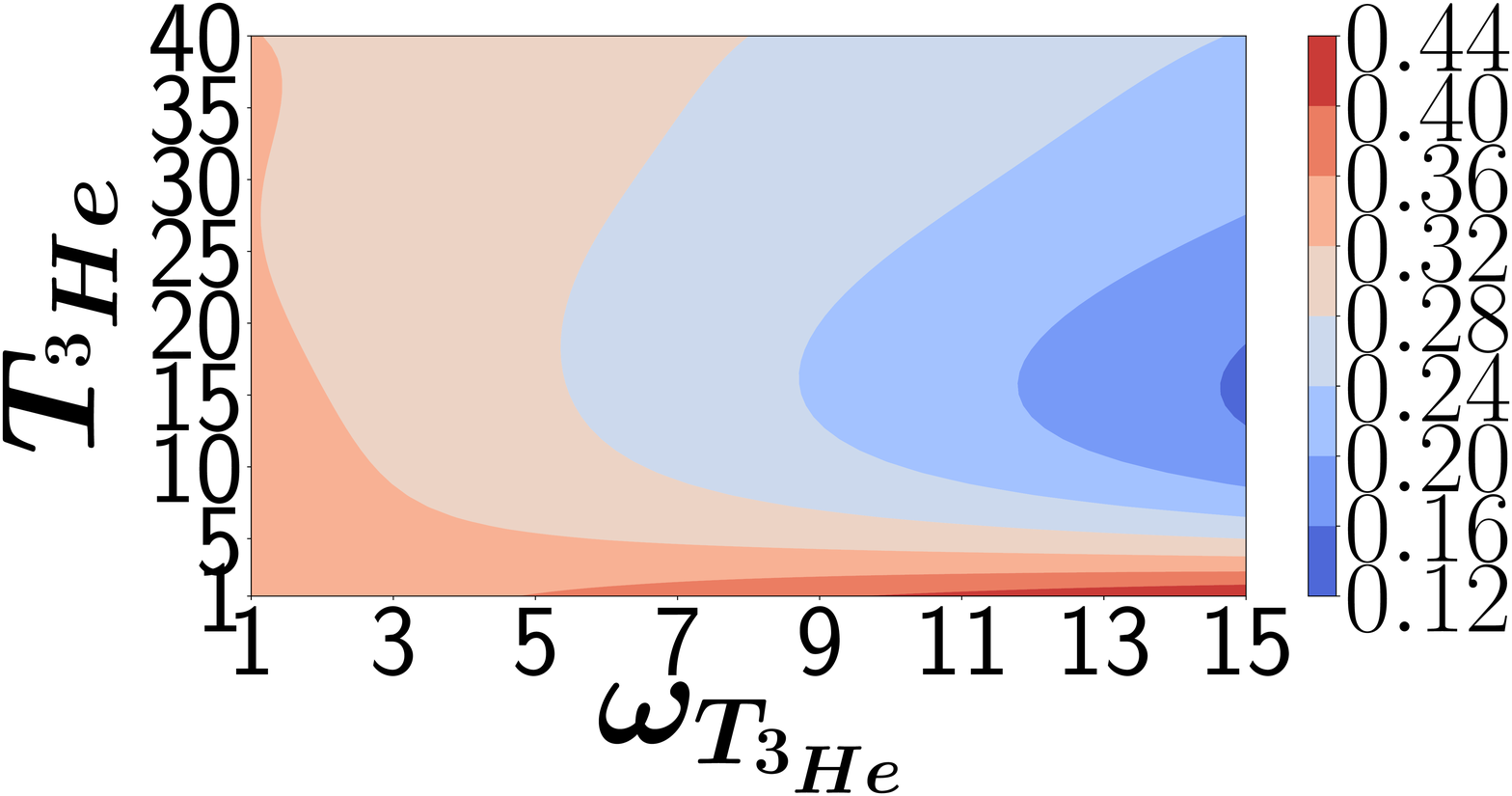}
\caption{$T_{^3He}$ and $\omega_{T_{^3He}}$}
\end{subfigure}
\begin{subfigure}{0.32\textwidth}
\includegraphics[width=1.0\textwidth]{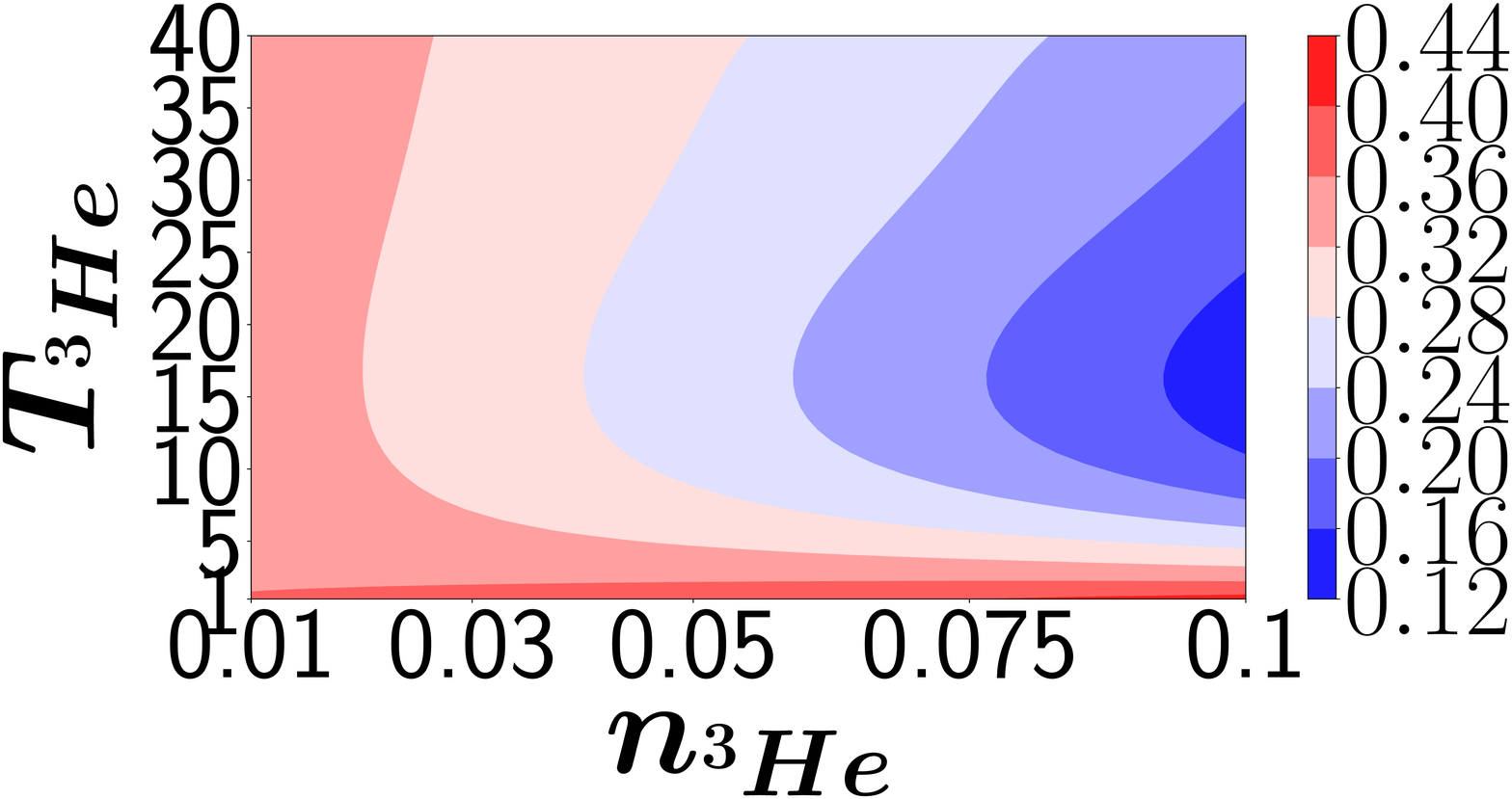}
\caption{$T_{^3He}$ and $n_{^3He}$}
\end{subfigure}
\begin{subfigure}{0.32\textwidth}
\includegraphics[width=1.0\textwidth]{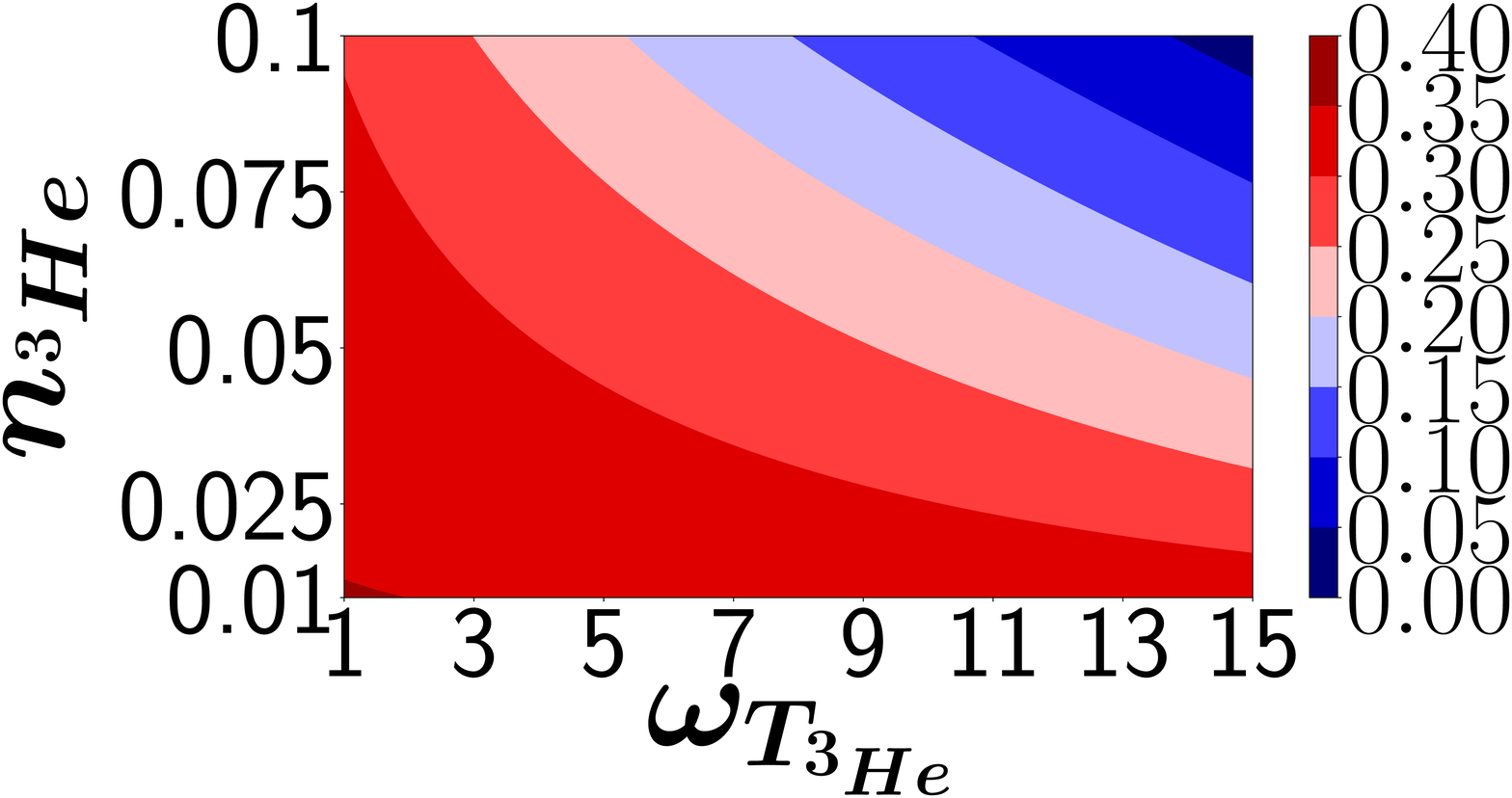}
\caption{$n_{^3He}$ and $\omega_{T_{^3He}}$}
\end{subfigure}
\caption{Contour plots of the most unstable linear growth rate, in units of $c_s / a$, at $k_y \rho_s = 0.5$ for different values of (a) $T_{^3He}$ and $\omega_{T_{^3He}}$, (b) $T_{^3He}$ and $n_{^3He}$ and (c) $n_{^3He}$ and $\omega_{T_{^3He}}$.}
\label{fig:interactions_2D}
\end{figure}

To further understand these two-parameter interactions, we depict, in Fig.~\ref{fig:interactions_2D}, the dependency of the ITG growth rate on each of the three combinations. 
By looking at Figure \ref{fig:interactions_2D}(a)--(b) we observe the characteristic ``sweet spot" in the $^3$He temperature, observed in Ref.~\cite{DiSiena_2018}, where the ITG growth rates exhibit a strong reduction, going from $\gamma [c_s / a] = 0.44$ to $\gamma [c_s / a] = 0.12$. 
Moreover, Fig.~\ref{fig:interactions_2D} shows that this linear ITG stabilization is proportional to the $^3$He logarithmic temperature gradient and its density.
The behaviour of the ITG growth rates with the $^3$He parameters is consistent with the numerical results and theoretical predictions of Ref.~\cite{DiSiena_PoP2019}. 
More precisely, Ref.~\cite{DiSiena_2018} proved that fast ions can interact with the ITG instability when the drift-frequency of the supra-thermal particles, $\omega_{df}$, gets close to the frequency of the underlying plasma micro-instabilities, $\omega$. 
This resonant condition is mainly controlled by the supra-thermal ion temperature $T_f$, being $\omega_{df}$ proportional to $T_f$. 
The minimum in the ITG growth rate in the $^3$He temperature is located exactly at $T_{^3He} = 12$, which, represents the ``optimal" value fulfilling the resonant condition $\omega = \omega_{df}$.

By means of gyrokinetic simulations and theory, it was shown in Refs.~\cite{DiSiena_2018,DiSiena_PoP2019} that the direction of this resonant energy exchange is determined by the drive term of the supra-thermal particles, and hence by their temperature and density logarithmic gradients. 
Fast ions are found to stabilize ITG modes only when $\omega_{T_{^3He}} \gg \omega_{n_{^3He}}$. 
This constraint is always fulfilled for the reference parameters employed throughout this paper, thus making the total sensitivity index of the $^3$He logarithmic density gradient negligible. 
This result is in agreement with the minor role played by the NBI fast deuterium, since this constraint on the logarithmic gradients is never fulfilled. 
In contrast, the role of the wave-particle resonant interaction in suppressing ITG modes is substantial for the ICRH $^3$He and increases with the $^3$He density and the logarithmic temperature gradient. 
The resonant energy exchange is indeed enhanced by these parameters, which, respectively, increase the $^3$He drive term and the $^3$He contribution in the field equations.

We conclude this section with perhaps the most striking results of the sensitivity-driven dimension-adaptive sparse grid approach: the sensitivity-driven dimension-adaptive approach required only $250$ {\sc Gene} simulations to complete the $21$ parameter scan. 
To put this number into perspective, a standard scan employing two points per parameter requires $2^{21} \approx 2.1 \times 10^6$ simulations in total. 
This remarkable result is a consequence of the small Sobol' indices -- as shown in Fig.~\ref{fig:sensitivity_test_case_3} -- which indicates that the sensitivity-driven algorithm did not need to explore many of the $21$ directions.
\section{Towards discharge optimization} \label{sec:nonlin_res}

The results of the previous section allow us to identify the main plasma parameters controlling the ITG growth rates for the wave-number $k_y \rho_s = 0.5$. 
In particular, we see that the Miller geometry parameters have total Sobol' indices close to zero.
In the present section, we restrict the number of free parameters to the first $14$ from Table \ref{tab:all_params} and fix the Miller geometry parameters to their reference values.

In Section \ref{subsec:lin_14D}, we consider a broad range of mode-numbers and ascertain the importance of the $14$ parameters by means of their total Sobol' indices. 
Moreover, we exploit the quasi-linear nature of the wave-particle mechanism and perform an optimization procedure to enhance the fast ion stabilization over a broad range of mode-numbers. 
This is done again by exploiting the sensitivity-driven approach, which, being based on interpolation, implicitly provides a surrogate for the linear flux-tube solver.
The relevance of this optimization procedure is demonstrated in Section \ref{subsec:nonlin_14D}, where nonlinear {\sc Gene} turbulence simulations are performed with both the reference and the optimized parameters, showing a significant reduction in the outward radial fluxes by more than two order of magnitude with the optimized setup.

\subsection{Finding a surrogate for the input-to-growth rate map and optimization procedure}
\label{subsec:lin_14D}
We employ the sensitivity-driven dimension-adaptive sparse grid approach to a broad range of mode-numbers to construct a model able to reproduce the growth rate function for each of the considered $k_y \rho_s$ and plasma parameters.
More precisely, we perform scans for $k_y \rho_s \in \{0.1, 0.2, 0.3, 0.4, 0.5, 0.6, 0.7, 0.8, 0.9\}$. 

\begin{figure}[htpb]
\centering
\includegraphics[width=1.0\textwidth]{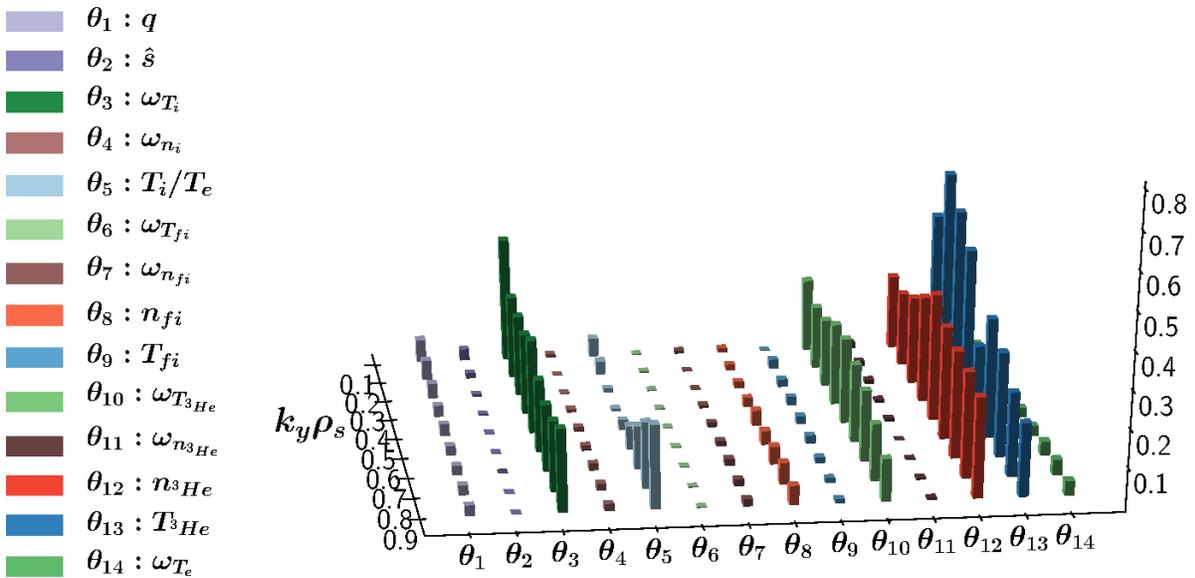}
\caption{Sensitivity information expressed in terms of total Sobol' indices for the scan involving $14$ parameters for $k_y \rho_s \in \{0.1, 0.2, 0.3, 0.4, 0.5, 0.6, 0.7, 0.8, 0.9\}$.}
\label{fig:sensitivity_test_case_2}
\end{figure}

We start by showing, in Fig.~\ref{fig:sensitivity_test_case_2}, the total sensitivity coefficients for all mode-numbers. 
Consistently with the results of Section \ref{sec:lin_res}, we observe that the most relevant parameters are the main ion logarithmic temperature gradient, the $^3$He temperature, its logarithmic gradient and the $^3$He density. 
However, as the mode-number increases, we also see a larger contribution of the ion-electron temperature ratio, $T_i/T_e$.

\begin{figure}[htpb]
\centering
\includegraphics[width=0.7\textwidth]{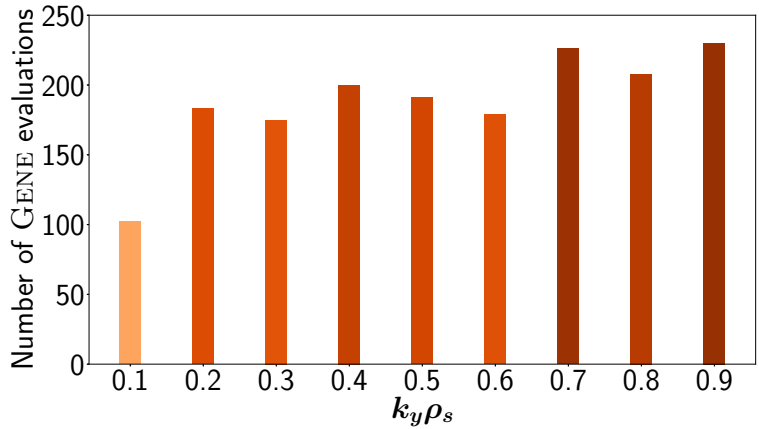}
\caption{Number of {\sc Gene} evaluations required by the sensitivity-driven dimension-adaptive sparse grid approach to perform the scan of Fig. \ref{fig:sensitivity_test_case_2}.}
\label{fig:cost}
\end{figure}

The number of {\sc Gene} evaluations necessary for the scans at all nine perpendicular wave-numbers is shown in Fig.~\ref{fig:cost}. 
We highlight that the sensitivity-driven approach requires at most $230$ {\sc Gene} simulations for the $14$ parameter scan. 
We once again see the tremendous computational savings that our sensitivity-driven strategy can bring to parameter scans. 
In scenarios in which one simulation is computationally very expensive, on present-day (or future) supercomputers, a total of $230$ simulations is not excessive.

Thus far, the sensitivity-driven dimension-adaptive approach was employed to compute the sensitivities of the parameters in the scan.
In the following, we exploit the fact that it also provides an interpolation approximation of the parameters-to-growth rate mapping and employ it as a surrogate for {\sc Gene} in solving a $14$-dimensional optimization problem.
% To ascertain the accuracy of this mapping, we estimate the mean-squared error (MSE), \textcolor{amaranth}{root-mean squared error (RMSE) and the relative error (RE), computed as follows:
% \begin{subequations} \label{eq:errors}
% \begin{align}
% \mathrm{MSE}(\gamma, \hat{\gamma}) = \frac{1}{M} \sum_{n=1}^M \left(\gamma_n - \hat{\gamma}_n\right)^2 \\
% \mathrm{RMSE}(\gamma, \hat{\gamma}) = \sqrt{\left(\frac{1}{M} \sum_{n=1}^M \left(\gamma_n - \hat{\gamma}_n\right)^2 \right)} \\
% \mathrm{RE}(\gamma, \hat{\gamma}) = \left. \sqrt{\sum_{n=1}^M \left(\gamma_n - \hat{\gamma}_n\right)^2} \right/ \sqrt{\sum_{n=1}^M \gamma_n^ 2}
% \end{align}
% \end{subequations}}
To ascertain the accuracy of this mapping, we estimate the mean-squared error (MSE), root-mean squared error (RMSE) and the relative error (RE), computed as follows:
\begin{subequations} \label{eq:errors}
\begin{align}
\mathrm{MSE}(\gamma, \hat{\gamma}) = \frac{1}{M} \sum_{n=1}^M \left(\gamma_n - \hat{\gamma}_n\right)^2 \\
\mathrm{RMSE}(\gamma, \hat{\gamma}) = \sqrt{\left(\frac{1}{M} \sum_{n=1}^M \left(\gamma_n - \hat{\gamma}_n\right)^2 \right)} \\
\mathrm{RE}(\gamma, \hat{\gamma}) = \left. \sqrt{\sum_{n=1}^M \left(\gamma_n - \hat{\gamma}_n\right)^2} \right/ \sqrt{\sum_{n=1}^M \gamma_n^ 2}
\end{align}
\end{subequations}
between the growth rate $\gamma$ computed by {\sc Gene} and the growth rate estimated by the sensitivity-driven approach, $\hat{\gamma}$, at $M$ random samples drawn from the uniform distribution bounds equal to the parameter bounds displayed in Table \ref{tab:all_params}.
% For example, at $k_y \rho_s = 0.5$, using $M = 1000$ yields $\mathrm{MSE}(\gamma, \hat{\gamma}) = 6.4746 \times 10^{-4}, \textcolor{amaranth}{\mathrm{RMSE}(\gamma, \hat{\gamma}) = 2.5445 \times 10^{-2}}$ \textcolor{amaranth}{and} $\textcolor{amaranth}{\mathrm{RE}(\gamma, \hat{\gamma}) = 8.8756 \times 10^{-2}}$.
For example, at $k_y \rho_s = 0.5$, using $M = 1000$ yields $\mathrm{MSE}(\gamma, \hat{\gamma}) = 6.4746 \times 10^{-4}, \mathrm{RMSE}(\gamma, \hat{\gamma}) = 2.5445 \times 10^{-2}$ and $\mathrm{RE}(\gamma, \hat{\gamma}) = 8.8756 \times 10^{-2}$.
Since the {\sc Gene} growth rates were computed with three digits of accuracy, the values of these errors are sufficiently small for our purposes.
Therefore, we see that at $k_y \rho_s = 0.5$, the sensitivity-driven dimension-adaptive approach yielded both a detailed description of the sensitivities of the growth rate and a sufficiently accurate approximation of the input-to-growth rate map, at a cost of only $230$ {\sc Gene} simulations in total.

We employ the surrogate to find the parameter set that minimizes the average growth rate for wave numbers $k_y \rho_s \in \{0.1, 0.2, 0.3, 0.4, 0.5\}$, which are the most unstable mode numbers in linear simulations. 
To this end, we solve the following constrained minimization problem:
\begin{equation} \label{eq:minimizer}
\begin{aligned}
\underset{\boldsymbol{\theta} \in \mathbb{R}^{14}}{\mathrm{min} \ \ \ } &  \frac{1}{5} \sum_{n=1}^5 \hat{\gamma}_n(\boldsymbol{\theta}) \\
\mathrm{subject \ to \ }  \ &  a_j \leq \theta_j \leq b_j, \quad j = 1, 2, 4, \ldots, 14 \\
\ & \theta_{3} = 4.5640
\end{aligned}
\end{equation}
where $\hat{\gamma}_n(\boldsymbol{\theta}) := \mathcal{U}^{d}_{\mathcal{L}}[\gamma_n](\boldsymbol{\theta})$ represents the growth rate estimated using the sensitivity-driven sparse grid surrogate. The subscript $n = 1, 2, \ldots, 5$ indicates that we consider the five perpendicular wave-numbers $k_y \rho_s \in \{0.1, 0.2, 0.3, 0.4, 0.5\}$. 
The minimization bounds, $a_j, b_j$ for $j = 1, 2, 4, \ldots, 14$, are the ones listed in Section \ref{subsec:lin_14D} (see Table \ref{tab:all_params}). 
Moreover, we fix the logarithmic main ion temperature gradient to the reference value to keep a constant ITG drive term. 
Without this constraint the minimization procedure will lead to the trivial set of parameters with the smallest gradient.

\begin{figure}[htpb]
\centering
\includegraphics[width=0.7\textwidth]{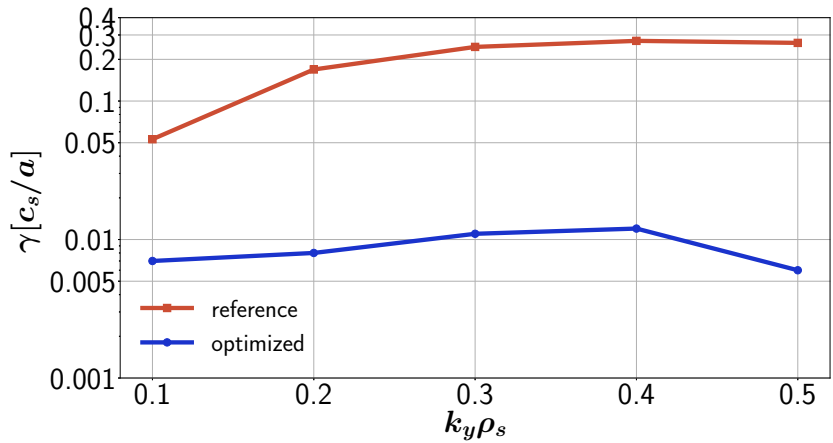}
\caption{Comparison of the growth rates -- in units of $c_s / a$ -- obtained using the minimizer in Table \ref{tab:minimizer} (blue line) and the reference parameters (orange line).}
\label{fig:minimizer}
\end{figure}

To perform the minimization \eqref{eq:minimizer}, we employ the Sequential Least Squares Programming (SLSQP) algorithm available in \texttt{python}'s optimization suite\footnote{\texttt{https://docs.scipy.org/doc/scipy/reference/optimize.minimize-slsqp.html}}. 
The solution is showed in Table \ref{tab:minimizer}, in the last column.
For an easier comparison, we also display the nominal values of the $14$ parameters in column three.
\begin{table}[htbp]
\centering
\begin{tabular}{|c|c|c|c|}
\hline
$\boldsymbol{\theta}$ & input parameter & nominal value & optimized value \\ 
\hline
$\theta_1$ & $q$ & $1.7364$ & $1.5148$ \\
$\theta_2$ & $\hat{s}$ & $0.5226$ & $0.5085$ \\
$\theta_3$ & $\omega_{T_i}$ & $4.5640$ & $4.5640$ \\
$\theta_4$ & $\omega_{n_i}$ & $0.0063$ & $0.0062$ \\
$\theta_5$ & $T_i/T_e$ & $0.9000$ & $1.0054$ \\
$\theta_6$ & $\omega_{T_{fi}}$ & $1.0323$ & $2.2708$ \\
$\theta_7$ & $\omega_{n_{fi}}$ & $4.7217$ & $5.1465$ \\
$\theta_8$ & $n_{fi}$ & $0.0600$ & $0.0541$ \\
$\theta_9$ & $T_{fi}$ & $9.8000$ & $22.4939$ \\
$\theta_{10}$ & $\omega_{T_{^{3}{He}}}$ & $7.4058$ & $14.5187$ \\
$\theta_{11}$ & $\omega_{n_{^{3}{He}}}$ & $0.5027$ & $1.5532$ \\
$\theta_{12}$ & $n_{^{3}{He}}$ & $0.0700$ & $0.0935$ \\
$\theta_{13}$ & $T_{^{3}{He}}$ & $6.9000$ & $19.0092$ \\
$\theta_{14}$ & $\omega_{T_{e}}$ & $2.2260$ & $2.0809$ \\
\hline
\end{tabular}
\caption{Solution for the constrained optimization problem \eqref{eq:minimizer}, displayed in the fourth column. For an easier comparison, we also show the nominal values of the $14$ parameters in the third column.}
\label{tab:minimizer}
\end{table}
In Fig.~\ref{fig:minimizer}, we show the growth rates obtained using the minimizer showed in Table \ref{tab:minimizer} for all five wave-numbers.
We also depict the growth rates obtained using the nominal values of the $14$ parameters.
We see that the minimizer leads to a significant stabilization: for $k_y \rho_s = 0.1$, the growth rate was reduced by about $86.7\%$, whereas for $k_y \rho_s = 0.5$, we obtained a reduction of around $97.7\%$.
\subsection{Nonlinear {\sc Gene} simulations: Confirming turbulence suppression by energetic particles}
\label{subsec:nonlin_14D}

To fully grasp the potential of the minimizer discussed in the previous section, we perform two nonlinear {\sc Gene} simulations: (i) for the reference parameters and (ii) for the optimized set of parameters showed in Table \ref{tab:minimizer}.

The radial box size is $170 \rho_s$ and the minimal $k_y \rho_s = 0.05$. 
We use $172$ points in the radial direction and $48$ points in the bi-normal direction.
We employed $24$ points along the field line.
Moreover, in velocity space, we use $32$ equidistant symmetric parallel velocity grid points and $16$ Gauss-Laguerre distributed magnetic moment points. 

\begin{figure}[htpb]
\centering
\includegraphics[width=0.7\textwidth]{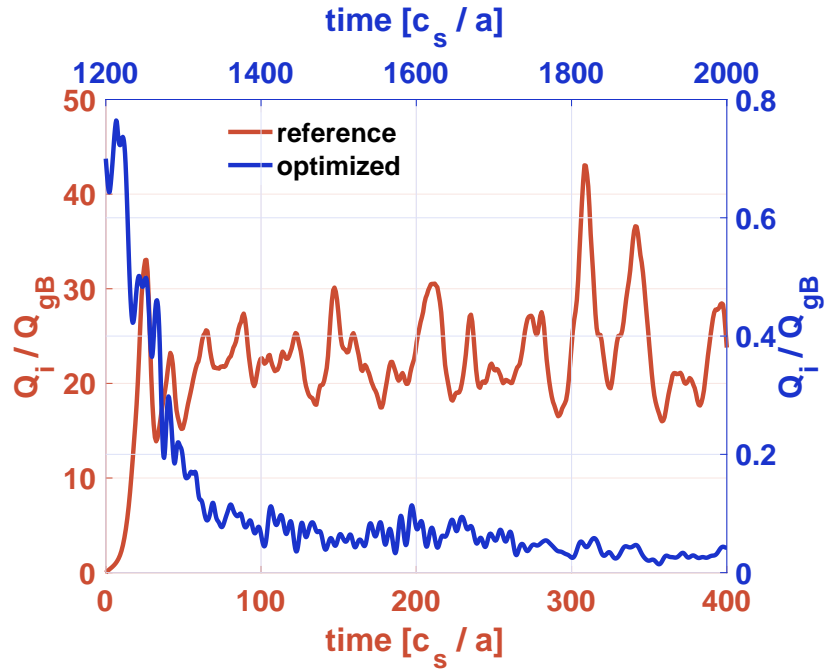}
\caption{Time trace on the main ion heat flux - in GyroBohm units - for the nominal (orange line) and optimized (blue line) set of parameters.}
\label{fig:nonlinear}
\end{figure}

The non-linear main ion heat fluxes -- in GyroBohm  units $Q_{gB} = T_e^{2.5}n_im_i^{0.5}/e^2B^2_0R^2_0$, with $e$ effective ion charge -- are illustrated in Fig.~\ref{fig:nonlinear} for the reference (orange line) and optimized (blue line) parameters.
A striking result is the overall turbulent stabilization achieved via the numerical optimization performed with the sensitivity-driven approach. 
The time-averaged ion heat flux in the saturated phase is almost totally suppressed with the optimized set of parameters. 
It is reduced by more than two order of magnitude, going from $Q_{i,gB}/Q_{gB} = 25$ to $Q_{i,gB}/Q_{gB} = 0.04$. 
The same stabilization is observed for the electron and fast ions turbulent fluxes (not shown here). 
These findings demonstrate the robustness of the sensitivity-driven optimizer and its high effectiveness in suppressing turbulent transport. 
It is worth mentioning here that the procedure employed throughout this paper is particularly successful in the nonlinear simulations due to the optimization of an underlying quasi-linear effect on turbulence, namely the wave-particle resonant interaction. 
A procedure involving an optimization of a nonlinear effect would require a sensitivity-driven approach surrogate built on nonlinear simulations. 
This will be the next natural step in this line of research.
\section{Conclusions} \label{sec:conclusions}

In the present paper, we have presented our newly developed sensitivity-driven dimension-adaptive sparse grid interpolation technique, which we have used to study the role of supra-thermal particles in suppressing ITG instabilities for a set of realistic JET-like parameters with both NBI fast deuterium and ICRH $^3$He.

The usefulness of the sensitivity-driven approach is two-fold.
First, it provides an efficient way of performing high-dimensional parameter scans, which is due to the adaptive refinement of the scanning grid, based on sensitivity information about the input parameter: if an input is not important, the algorithm will not invest effort in its corresponding direction.
At the end of the adaptive process, the sensitivity-driven approach provides a detailed description about the sensitivity of each input parameter, including all their interactions.
% Moreover, it also yields the expectation and standard deviation of the output of interest.
The second useful feature of the approach is that it provides an approximation of the parameter-to-output map, which can be used in further computationally expensive tasks, such as high-dimensional optimization.

We showcased the efficiency of the sensitivity-driven approach in a $21$ parameter scan for which only $250$ simulations where necessary.
We also considered a subset of $14$ parameters and used the approach to find an approximation of the parameter-to-growth rate map at nine bi-normal wave-numbers.
For this task, we needed at most $230$ simulations per wave-number.
We used this map to minimize the average growth rate at the five most unstable wave-numbers in the nonlinear simulations.
The minimizer lead to a significant reduction of the growth rate -- up to $97.7\%$ from the nominal value. 
We further showcased this stabilization by performing two non-linear simulations, comparing the reference results with the results yielded by our minimizer, which showed a significant reduction of the non-linear heat fluxes by more than two order of magnitude.

In summary, the sensitivity-driven dimension-adaptive sparse grid approach can be used to very efficiently perform scans in high-dimensional parameter spaces, saving up to several orders of magnitude in computational effort with respect to conventional scanning methods. This implies that certain optimization procedures, involving large numbers of computations and deemed out of reach, become feasible now. This includes, in particular, transport studies based on (quasi-)linear and/or non-linear gyrokinetic simulations, a task that is essential to improve plasma confinement. 
Moreover, this approach can easily be generalized to many other problems in fusion research, and well beyond. Possible applications include the optimization of the performance of existing devices as well as the design of new ones. Machines like ITER, DEMO, and fully optimized stellarators are expected to provide many fruitful application areas.
\section*{Acknowledgements}
Numerical simulations were performed at the MARCONI-Fusion supercomputer at CINECA, Italy.
We also acknowledge the useful comments and suggestions from the anonymous referees. 

\section*{References}
\bibliographystyle{iopart-num}
\bibliography{paper_UQ_FSJ}

\end{document}